\documentstyle[12pt]{article}
\documentstyle{fleqn}

\begin{document}
\begin{flushright}
OU-HET/185\\hep-th/9312151\\December 1993
\end{flushright}
\vspace{0.5in}
\begin{center}\Large{\bf Reduced Phase Space of the first order \\
Einstein Gravity on ${\bf R}\times T^{2}$}\\
\vspace{1cm}\renewcommand{\thefootnote}{\fnsymbol{footnote}}
\normalsize\ Kiyoshi Ezawa\footnote[1]{e-mail address:
 ezawa@oskth.kek.jp}\setcounter{footnote}{0}

\vspace{0.5in}

        Department of Physics \\
        Osaka University, Toyonaka, Osaka 560, Japan\\
\vspace{0.1in}
\end{center}
\vspace{1.5in}
\baselineskip 17pt
\begin{abstract}
Chern-Simons formulation of the 2+1 dimensional Einstein gravity
with negative cosmological constant is investigated when the
spacetime has the topology ${\bf R}\times T^{2}$. The physical
phase space is shown to be a direct product of two sub-phase
spaces having symplectic structures with opposite signs.
Each sub-phase space is found to be
a non-Hausdorff manifold plus a set of
measure zero. Geometrical interpretation of each point in the
phase space is also given. A prescription to quantize this phase
space is proposed.
\end{abstract}
\newpage

\baselineskip 20pt

 \ \ \ \ Since the first order formalism  of 2+1 dimensional
Einstein gravity was shown to be equivalent to the Chern-Simons
gauge theories with
noncompact gauge groups\cite{witte}\cite{gold},
many works have appeared on this \lq\lq Chern-Simons gravity"(CSG).
Particularly in the case where the spacetime topology
is ${\bf R}\times T^{2}$ and the cosmological constant vanishes,
various aspects of the CSG including its geometrical interpretation
and the structure of its phase space seem to have been elucidated
\cite{carli} \cite{ander} \cite{carl2} \cite{louko} \cite{unruh}.

As for the case with nonvanishing cosmological constant, except
a series of works on the holonomy algebra which are made by
Nelson and Regge \cite{nelson}\cite{CN}, relatively few people
deal with this case\cite{cenalo}.
In the previous paper\cite{ezawa}, we have
shown that the physical phase space in the negative cosmological
constant case has nine sectors when the spacetime has the topology
${\bf R}\times T^{2}$ and one of these sectors is in 1 to 2
correspondence with the ADM phase space. However we knew
little about the remaining eight sectors.
In this paper, we will give the topological and symplectic
structures to this phase space as a whole.
We will find that this phase space is not equipped with
a cotangent bundle structure.
We will give geometrical interpretations to each of the nine
sectors.
We will also discuss its quantization.

Before going into the subject, we make some remarks.
We use the notation in the previous paper\cite{ezawa}.
In giving the geometrical meaning to each sector of the phase
space, we will use the construction of spacetimes proposed by
Witten\cite{witte}. Namely we identify the spacetime $M$ with
a quotient space ${\cal F}/G$, where ${\cal F}$ is a subspace of
the anti-de Sitter space $AdS^{3}$ and $G$ is a subgroup of
$SO(2,2)$ which is specified by
a point on the physical phase space.
Since we want to associate a spacetime with every point on the
phase space, we will tacitly take the universal covering of
${\cal F}$ if
$G$ involves rotations. It therefore seems to be natural to use
as the gauge group the universal covering group
$\widetilde{SO}(2,2)$ of $SO(2,2)$. Thus a $2\pi$-rotation is no
longer equivalent to the identity. For simplicity of the
analysis, we will use its subgroup
$\widetilde{SO}_{0}(2,2)$ which is
connected with the identity. The prescription
for reducing the phase space used in ref.\cite{ezawa} still remains
valid also in this case (though with a few exceptions),
since the conjugation transformation
by an element of $\widetilde{SO}(2,2)$ induces the adjoint
representation of the $SO(2,2)$ Lie algebra, whose group structure
is $SO(2,2)$.

First we look into the topological structure of the physical
phase space. We start with the reduced action of the CSG with the
negative cosmological constant $\Lambda=-L^{-2}$ on ${\bf R}\times
\Sigma$
\begin{equation}
I^{\ast}_{W}=\frac{L}{2}\int_{\bf R}dt\int_{\Sigma}\eta_{ab}
\{-^{(2)}A^{(+)a}\wedge^{\mbox{ }(2)}\dot{A}^{(+)b}+^{(2)}
A^{(-)a}\wedge^{\mbox{ }(2)}\dot{A}^{(-)b}\}.\label{eq:action}
\end{equation}
We have used (the pullback into $\Sigma$ of) the (anti-)self-dual
$SO(2,2)$ connection
\begin{equation}
^{(2)}A^{(\pm)a}\equiv (\frac{1}{2}\epsilon^{a}_{\mbox{ }bc}
\omega^{bc}_{i}\pm
\frac{1}{L}e^{a}_{i})dx^{i}, \label{eq:sdconn}
\end{equation}
where $e^{a}_{\mu}$ and $\omega^{ab}_{\mu}$ are a triad and a spin
connection respectively. Note that $^{(2)}A^{(\pm)a}$ is flat since
we have solved the constraint. We see from the action
(\ref{eq:action}) that the physical phase space ${\cal M}$
is expressed as a direct product:
$$
{\cal M}={\cal M}^{(+)}\times{\cal M}^{(-)},
$$
where ${\cal M}^{(\pm)}$ is the sub-phase space which is the moduli
space of flat (anti-)self-dual $SO(2,2)$ connections modulo gauge
transformations. We can also realize that ${\cal M}^{(+)}$
and ${\cal M}^{(-)}$ have symplectic structures with
opposite signs.

Let us now consider the case where $\Sigma\approx T^{2}$.
We will denote two generators of $\pi_{1}(T^{2})$ by $\alpha$ and
$\beta$. We know that each sub-phase spaces ${\cal M}^{(\pm)}$
consists of three subsectors ${\cal M}^{(\pm)}_{S}$,
${\cal M}^{(\pm)}_{N}$, and ${\cal M^{(\pm)}}_{T}$ \cite{ezawa}
(plus a set ${\cal M}^{(\pm)}_{0}$ with measure zero\footnote{
The ${\cal M}^{(\pm)}_{0}$ is the union of a point
$\{e^{\pm}\}\equiv\{S^{(\pm)}[\alpha]=S^{(\pm)}[\beta]=1\}$ and
four countable sets of 1-parameter families:
\begin{eqnarray*}
{\cal M^{(\pm)}}_{\bullet\pm}\equiv
\bigcup_{n\in{\bf Z}\backslash\{0\}}{\cal M}^{(\pm)}_{n,\pm}&,&
{\cal M}^{(\pm)}_{n,\pm}\equiv\left\{S^{(\pm)}[\alpha]=\exp(
2n\pi\lambda_{0}),\mbox{ }S^{(\pm)}[\beta]=\exp(
\pm\lambda_{0}+b\lambda_{2})\mbox{ }|\mbox{ } b\geq 1 \right\}, \\
{\cal M^{(\pm)}}_{\pm\bullet}\equiv
\bigcup_{m\in{\bf Z}\backslash\{0\}}{\cal M}^{(\pm)}_{\pm,m}&,&
{\cal M}^{(\pm)}_{\pm,m}\equiv\left\{S^{(\pm)}[\alpha]=\exp(\pm
\lambda_{0}+a\lambda_{2}),\mbox{ }S^{(\pm)}[\beta]=\exp(2m\pi
\lambda_{0})\mbox{ }|\mbox{ } a\geq 1 \right\}.
\end{eqnarray*}  } ).
The ${\cal M}^{(\pm)}_{S}$ is
parametrized by\footnote{$\lambda_{a}$ denotes (the universal
covering version of) the pseudo-Pauli
matrices.}
\begin{equation}
S^{(\pm)}[\alpha]=\exp(\lambda_{2}\alpha_{\pm}),\quad
S^{(\pm)}[\beta]=\exp(\lambda_{2}\beta_{\pm}), \label{eq:ssec}
\end{equation}
with $(\alpha_{\pm},\beta_{\pm})\in ({\bf R}^{2}\backslash\{(0,0)\})
/{\bf Z}_{2}$.\footnote{${\bf Z}_{2}$ in the denominator is
generated by the internal inversion:
$(\alpha_{\pm},\beta_{\pm})\rightarrow
-(\alpha_{\pm},\beta_{\pm})$.}
Its symplectic structure is given by
\begin{equation}
\mp Ld\alpha_{\pm}\wedge d\beta_{\pm}.\label{eq:ssymp}
\end{equation}
Parametrization of the ${\cal M}^{(\pm)}_{N}$ is\footnote{
This parametrization is different from that in \cite{ezawa}.
In fact the former includes the latter as a spesial case with $
\theta_{\pm}\in(-\pi/2,\pi/2)$. }
\begin{equation}
S^{(\pm)}[\alpha]=
\exp\{(\lambda_{0}\pm\lambda_{2})\cos\theta_{\pm}\},\quad
S^{(\pm)}[\beta]=\exp\{
(\lambda_{0}\pm\lambda_{2})\sin\theta_{\pm}\},\label{eq:nsec}
\end{equation}
with $\theta_{\pm}+2\pi$ being identified with $\theta_{\pm}$.
This sector by itself does not have a symplectic structure.
The ${\cal M}^{(\pm)}_{T}$ is expressed by the following
parametrization
\begin{equation}
S^{(\pm)}[\alpha]=\exp(\lambda_{0}\rho_{\pm}),\quad
S^{(\pm)}[\beta]=\exp(\lambda_{0}\sigma_{\pm}),\label{eq:tsec}
\end{equation}
with $(\rho_{\pm},\sigma_{\pm})\in{\bf R}^{2}\backslash\{(0,0)\}$.
Its symplectic structure is:
\begin{equation}
\pm Ld\rho_{\pm}\wedge d\sigma_{\pm}.\label{eq:tsymp}
\end{equation}

We would like to provide a construction in which these three
subsectors ${\cal M}^{(\pm)}_{S}$,
${\cal M}^{(\pm)}_{N}$ and ${\cal M}^{(\pm)}_{T}$ appear in one
parametrization. It turns out that this unification can
be done as in the $\Lambda=0$ case\cite{louko}.
For this purpose we first consider two commuting
$\widetilde{PSL}(2,{\bf R})$ holonomies in the following form:
\begin{eqnarray}
S^{(\pm)}[\alpha]&=& \exp\left[\cos\theta_{\pm}
\left\{\left(r_{\pm}+\sqrt{r_{\pm}^{\mbox{ }2}+1}\right)^{1/2}
\lambda_{0}\pm\left(-r_{\pm}+\sqrt{r_{\pm}^{\mbox{ }2}+1}\right)
^{1/2}\lambda_{2}\right\}\right], \nonumber \\*
S^{(\pm)}[\beta]&=& \exp\left[\sin\theta_{\pm}
\left\{\left(r_{\pm}+\sqrt{r_{\pm}^{\mbox{ }2}+1}\right)^{1/2}
\lambda_{0}\pm\left(-r_{\pm}+\sqrt{r_{\pm}^{\mbox{ }2}+1}\right)
^{1/2}\lambda_{2}\right\}\right]. \label{eq:unify}
\end{eqnarray}
The above holonomies with $r_{\pm}<0$, $r_{\pm}=0$ and
$r_{\pm}>0$ give parametrization of ${\cal M}_{S}$,
${\cal M}_{N}$ and ${\cal M}_{T}$ respectively.
Relations between these new parameters $(r_{\pm},\theta_{\pm})$
and the old ones $(\alpha_{\pm},\beta_{\pm})$ for
${\cal M}^{(\pm)}_{S}$ and $(\rho_{\pm},\sigma_{\pm})$ for
${\cal M}^{(\pm)}_{T}$ are obtained by performing on
(\ref{eq:unify}) the conjugation using $\exp(
\mp\lambda_{1}\Phi_{\pm})$ with $\Phi_{\pm}=\frac{1}{2}
\ln\{|r_{\pm}|/(\sqrt{r_{\pm}^{\mbox{ }2}+1}+1)\}$:
\begin{eqnarray}
(\alpha_{\pm},\beta_{\pm})=\pm\sqrt{-2r_{\pm}}
(\cos\theta_{\pm},\sin\theta_{\pm})\quad for
\quad r_{\pm}<0, \label{eq:suni} \\
(\rho_{\pm},\sigma_{\pm})=\quad \sqrt{2r_{\pm}}
(\cos\theta_{\pm},\sin\theta_{\pm})\quad for
\quad r_{\pm}>0. \label{eq:tuni}
\end{eqnarray}
Using the new parametrization, symplectic structures
(\ref{eq:ssymp})
(\ref{eq:tsymp}) are expressed by the unified form:
\begin{equation}
\pm Ldr_{\pm}\wedge d\theta_{\pm}. \label{eq:Symp}
\end{equation}
Besides, vanishing of the symplectic structure in
${\cal M}^{(\pm)}_{N}$ can be explained by the fact that
$r_{\pm}$ is a constant ({\it i.e.} zero) in this subsector.

In summary, we give the topological structure of
$$
{\cal M}^{\prime(\pm)}\equiv{\cal M}^{(\pm)}\backslash{\cal M}
^{(\pm)}_{0} ={\cal M}^{(\pm)}_{S}\cup{\cal M}^{(\pm)}_{N}\cup
{\cal M}^{(\pm)}_{T}.
$$
We should notice that the period of the parameter $\theta_{\pm}$
is $\pi$ for $r_{\pm}<0$ and $2\pi$ for $r_{\pm}\geq 0$. The
${\cal M}^{\prime(\pm)}$ defined above therefore turns out to be
a non-Hausdorff manifold constructed by gluing together a punctured
cone (the ${\cal M}^{(\pm)}_{S}$) and a punctured plane
(the ${\cal M}^{(\pm)}_{T}$) at the puncture in the one to two
fashion. The circle which serves as the glue is provided by the
${\cal M}^{(\pm)}_{N}$. The structure near the puncture precisely
coincides with that of the base space of cotangent bundle structure
of the phase space in the case without
cosmological constant \cite{louko}. In the case with
negative cosmological constant, however, the phase space
${\cal M}$ does not have a cotangent bundle structure even after
the removal of the set involving ${\cal M}^{(\pm)}_{0}$.
The phase space is represented by the direct product of two
non-Hausdorff manifolds plus
the set of measure zero.

Next we construct a spacetime from each point in the phase space.
Henceforth we will use $(x,y)$ as periodic coordinates on $T^{2}$
with period $1$. Identification conditions are therefore obvious.
We only consider the
subspace ${\cal M}^{\prime}\equiv{\cal M}^{\prime(+)}\times
{\cal M}^{\prime(-)}$ with nonzero measure,
which consists of the nine sectors. We will
denote these sectors as ${\cal M}_{(\Psi,\Phi)}\equiv
{\cal M}^{(+)}_{\Psi}\times{\cal M}^{(-)}_{\Phi}$ ($\Psi,\Phi
=S,N,T$).

As an illustration we review the spacetime construction
from ${\cal M}_{(S,S)}$ \cite{ezawa}. The simplest connection which
gives the holonomies (\ref{eq:ssec}) is
\begin{equation}
A^{(\pm)}\equiv\lambda_{a}A^{(\pm)a}_{\mu}dx^{\mu}
=\lambda_{2}d\varphi_{\pm}, \quad
(\varphi_{\pm}\equiv\alpha_{\pm}x+\beta_{\pm}y).
\end{equation}
By performing the time-dependent gauge transformation
$g^{(\pm)}=e^{\mp\lambda_{0}t}$ and by extracting the triad part,
we can construct the spacetime metric
\begin{equation}
L^{-2}ds^{2}=-dt^{2}+\cos^{2}t\mbox{ }d\left(
\frac{\varphi_{+}-\varphi_{-}}{2}\right)^{2}+
\sin^{2}t\mbox{ }d\left(\frac{\varphi_{+}+\varphi_{-}}{2}
\right)^{2}.     \label{eq:SS}
\end{equation}
Parametrization of the $AdS^{3}$ which reproduces this metric is:
\begin{equation}\begin{array}{ll}
(T,X,Y,Z)=L(\sin t\cosh\frac{\varphi_{+}+\varphi_{-}}{2},&
\sin t\sinh\frac{\varphi_{+}+\varphi_{-}}{2}, \\
&\cos t\sinh\frac{\varphi_{+}-\varphi_{-}}{2},
\cos t\cosh\frac{\varphi_{+}-\varphi_{-}}{2}). \end{array}
\label{eq:SSP}
\end{equation}
We should  remark that the periodicity condition for the
above parametrization is expressed by the identification under two
$\widetilde{SO}_{0}(2,2)$ transformations given by the holonomies
(\ref{eq:ssec}). Witten's construction therefore seems to be
equivalent to the standard construction explained above.

Indeed, it turns out that these two alternative constructions
give the same spacetime also to the remaining eight sectors.
We will omit the detail of its derivation and give only
parametrization in the $AdS^{3}$ which represent the
spacetime constructed from a point in each sectors.\footnote{
We always consider that
$$T^{2}-X^{2}-Y^{2}+Z^{2}=L^{2}$$
holds, and assume that the universal covering is taken
if necessary. The metric is obtained by substituting the
parametrization into the pseudo-Minkowski metric:
$$
ds^{2}=-dT^{2}+dX^{2}+dY^{2}-dZ^{2}.
$$}$^{,}$
\footnote{We define the following new coordinates on $T^{2}$:
$$
\eta_{\pm}\equiv x\cos\theta_{\pm}+y\sin\theta_{\pm},\quad
r_{\pm}\equiv\rho_{\pm}x+\sigma_{\pm}y.
$$}\\
${\cal M}_{(N,N)}$:
\begin{equation}\begin{array}{l}
X+Z=Le^{t}, \quad (T,Y)=Le^{t}\left(
\frac{\eta_{+}-\eta_{-}}{2},\frac{\eta_{+}+\eta_{-}}{2}\right).
\end{array}\label{eq:NN}
\end{equation}
${\cal M}_{(T,T)}$:
\begin{equation}
\begin{array}{ll}
L^{-1}(T,X,Y,Z)=(\cosh t\cos\frac{r_{+}-r_{-}}{2},&\sinh t\cos
\frac{r_{+}+r_{-}}{2},  \\
& \sinh t\sin\frac{r_{+}+r_{-}}{2},
-\cosh t\sin\frac{r_{+}-r_{-}}{2}).
\end{array}\label{eq:TT}
\end{equation}
${\cal M}_{(T,S)}$:
\begin{equation}
\begin{array}{l}
L^{-1}(T,X,Y,Z)=   \\
\qquad \cosh t(\cosh\frac{\varphi_{-}}{2}\cos\frac{r_{+}}{2},
\sinh\frac{\varphi_{-}}{2}\cos\frac{r_{+}}{2},
\sinh\frac{\varphi_{-}}{2}\sin\frac{r_{+}}{2},
-\cosh\frac{\varphi_{-}}{2}\sin\frac{r_{+}}{2})\\
\qquad  +\sinh t(\sinh\frac{\varphi_{-}}{2}\sin\frac{r_{+}}{2},
\cosh\frac{\varphi_{-}}{2}\sin\frac{r_{+}}{2},
-\cosh\frac{\varphi_{-}}{2}\cos\frac{r_{+}}{2},
\sinh\frac{\varphi_{-}}{2}\cos\frac{r_{+}}{2}).
\end{array}\label{eq:TS}
\end{equation}
${\cal M}_{(S,N)}$:
\begin{equation}
\begin{array}{l}
L^{-1}(T,Y)=(\sin t\cosh\frac{\varphi_{+}}{2},
\cos t\sinh\frac{\varphi_{+}}{2})+\frac{\eta_{-}
(\cos t\cosh\frac{\varphi_{+}}{2}
+\sin t\sinh\frac{\varphi_{+}}{2})}{2}(-1,1),\\
			         \\
L^{-1}(Z,X)=(\cos t\cosh\frac{\varphi_{+}}{2},
\sin t\sinh\frac{\varphi_{+}}{2})-\frac{\eta_{-}
(\cos t\sinh\frac{\varphi_{+}}{2}
+\sin t\cosh\frac{\varphi_{+}}{2})}{2}(-1,1).
\end{array}\label{eq:SN}
\end{equation}
${\cal M}_{(T,N)}$:
\begin{equation}\begin{array}{l}
L^{-1}(T,Y)=(\cosh t\cos\frac{r_{+}}{2},\sinh t\sin\frac{r_{+}}{2})
+\frac{\eta_{-}(\sinh t\cos\frac{r_{+}}{2}
-\cosh t\sin\frac{r_{+}}{2})}{2}(-1,1), \\
                        \\
L^{-1}(Z,X)=(-\cosh t\sin\frac{r_{+}}{2},\sinh t\cos\frac{r_{+}}{2})
-\frac{\eta_{-}(\cosh t\cos\frac{r_{+}}{2}
+\sinh t\sin\frac{r_{+}}{2})}{2}(-1,1).
\end{array}\label{eq:TN}
\end{equation}
As for the other three sectors ${\cal M}_{(S,T)}$,
${\cal M}_{(N,S)}$ and ${\cal M}_{(N,T)}$, the following holds
generically. The metric obtained from a point in
${\cal M}_{(\Phi,\Psi)}$ ($\Phi\neq\Psi$) can be made into the same
form as the one obtained from ${\cal M}_{(\Psi,\Phi)}$ with the
subscripts $\pm$ replaced by $\mp$. On the other hand, the triad
and the parametrization in the former are respectively obtained by
reversing the orientation of the triad and by replacing
$Z$ with $-Z$ in the latter.
Taking these facts into account, we can
say that the universe obtained from a point in
${\cal M}_{(\Phi,\Psi)}$ is the \lq\lq mirror image" of that in
${\cal M}_{(\Psi,\Phi)}$.

The eight sectors except ${\cal M}_{(S,S)}$ give spacetimes in
which each torus $T^{2}$ is timelike, so they do not correspond to
the ordinary ADM formalism. Though timelike closed curves contained
in these tori are forbidden
if we require the causality, there seems
to be no reason to suppress them in the quantum gravity, which is
the quantum theory of spacetime. So we can consider
that each point in ${\cal M}^{\prime}\backslash{\cal M}_{(S,S)}$
gives such an \lq\lq exotic" spacetime \cite{louko}.

We know that the ${\cal M}_{(S,S)}$ is in 1 to 2 correspondence
with the ADM formalism \cite{ezawa}. The expressions of the ADM
variables (complex modulus $m$, conjugate momentum $p$ and
Hamiltonian $H$) in terms of new parameters $(r_{\pm},\theta_{\pm})
$ are as follows:
\begin{eqnarray}
m&=&\frac{e^{it}\sin\theta_{+}\sqrt{-2r_{+}}
+e^{-it}\sin\theta_{-}\sqrt{-2r_{-}}}
{e^{it}\cos\theta_{+}\sqrt{-2r_{+}}
+e^{-it}\cos\theta_{-}\sqrt{-2r_{-}} }\quad, \\
p&=&\frac{-iL}{4\sin t\cos t}\left(e^{-it}\cos\theta_{+}
\sqrt{-2r_{+}}+e^{it}\cos\theta_{-}\sqrt{-2r_{-}}\right)^{2}, \\
H&=&\frac{-L}{\sin t\cos t}\sin(\theta_{+}-\theta_{-})
\sqrt{r_{+}r_{-}}\quad,
\end{eqnarray}
which are essentially the same as those given in ref.\cite{CN}.
These new parameters are related with the parameters $(\alpha,
\beta,u,v)$ in \cite{ezawa} by an ordinary canonical transformation
\begin{eqnarray}
2L(vd\alpha-ud\beta)&=&L(r_{+}d\theta_{+}-r_{-}d\theta_{-})+dV,
\nonumber \\
V(\alpha,\beta,\theta_{+},\theta_{-})&=&
-2L
\frac{(\alpha\sin\theta_{-}-\beta\cos\theta_{-})
(\alpha\sin\theta_{+}-\beta\cos\theta_{+})}
{\sin(\theta_{+}-\theta_{-})}\quad .\label{eq:cannon}
\end{eqnarray}
However, the canonical transformation from the ADM variables to
the new parameters is {\it singular} in the sense that it does
not contain the generating function:
\begin{equation}
{\rm Re}(\bar{p}dm)-Hdt=L(r_{+}d\theta_{+}-r_{-}d\theta_{-}).
\label{eq:can2}
\end{equation}
We conjecture that this singular nature is related to the
fact that $r_{\pm}$ and therefore $p$ cannot be expressed in terms
of $(m,\bar{m},\theta_{+},\theta_{-})$ alone.

We should note that the spacetimes we have given are not the
unique ones constructed from the points in ${\cal M}^{\prime}$.
It is because the gauge group $SO(2,2)$ (or $\widetilde{SO}_{0}
(2,2)$) is in fact larger than the direct product of the 2+1
dimensional local Lorentz group and the group of diffeomorphisms
\cite{louko}\cite{unruh}.
For illustration, we consider the ${\cal M}_{(S,S)}$. By choosing
time dependent gauge transformations other than that giving the
spacetime (\ref{eq:SS}), we can construct various spacetimes.
There are for example Louko-Marolf-type universe \cite{louko} in
which the evolution of the torus is roughly \footnote{
The notation $T\times S$ for example denotes the torus with
$(T,X)$- and $(Z,Y)$-directions being respectively timelike
and spacelike.}
$$T\times S\rightarrow S\times S\rightarrow S\times T, $$
and Unruh-Newbury-type universe in which the rough evolution of
the torus is:
$$
S\times T\rightarrow S\times S\rightarrow T\times S\rightarrow
S\times S\rightarrow S\times T,
$$
with an intermediate singularity in $T\times S$.
Though these spacetimes coincide with one another in the region
where the ADM is
well-defined ($T>|X|$, $Z>|Y|$), their behaviors in the other
region vary considerably by the choice of gauge.
There seems to be no
criterion for choosing the most relevant gauge.

Finally we consider how to quantize the subspace ${\cal M}^{\prime}
$ of the physical phase space. Since ${\cal M}^{\prime}$ does not
have a cotangent bundle structure, we cannot naively perform the
quantization where quantum states are represented by functions
of {\it coordinates}. We have to invoke geometric quantization
scheme \cite{wood} such as Kahler quantization\cite{witte}. To many
physicists these methods are not familiar. By a trick we provide
the ${\cal M}^{\prime}$ with a cotangent bundle structure.

First we consider the symmetry under large
diffeomorphisms, in particular the inversion:
\begin{equation}
I:\mbox{ }(\alpha,\beta)\longrightarrow -(\alpha,\beta)
\qquad \left(or\quad(x,y)\rightarrow -(x,y)\right),
\end{equation}
which induces the following {\em simultaneous} transformations:
\begin{equation}
I:\mbox{ }(\theta_{\pm},r_{\pm})\longrightarrow
(\theta_{\pm}+\pi,r_{\pm}). \label{eq:inver}
\end{equation}
${\cal M}^{\prime}$ does not have a cotangent bundle
structure even after imposing this symmetry. If we adopt one of the
following {\em artificial} prescription, however,
the resulting phase space
${\cal M}^{\circ}$ acquires a cotangent bundle structure
${\cal M}^{\circ}={\bf T}_{\ast}{\cal B}$ with the base space
${\cal B}$:\\
i) We impose the inversion symmetry {\em independently} on
${\cal M}^{\prime(+)}$ and on ${\cal M}^{\prime(-)}$. Then
${\cal M}^{\circ}={\bf T}_{\ast}T^{2}$. The coordinates $
(\theta_{+},\theta_{-})$ which parametrize the base space $T^{2}$
have periods $\pi$. \\
ii) We assume that $(\theta_{+},\theta_{-}+\pi)$ can be
distinguished from $(\theta_{+},\theta_{-})$ {\em even when either
$r_{+}$ or $r_{-}$ is negative}. This involves the
assumption that the ${\cal M}_{(S,S)}$ is not in 1 to 2
correspondence but equivalent with the ADM phase space.
The base space ${\cal B}$ then becomes a torus
$T^{2}$ different from that in i).

If we use one of these cotangent
bundle structures, we can construct
a representation where the quantum states are functions of $
(\theta_{+},\theta_{-})$. In the quantum theory the canonical
variables $(\theta_{\pm},r_{\pm})$ are promoted to the basic
operators which satisfy the canonical commutation relations derived
from the symplectic structure (\ref{eq:Symp}):
\begin{equation}
[\hat{\theta}_{\pm},\hat{r}_{\pm}]=\pm i\frac{1}{L}\quad, \quad
\mbox{\it zero otherwise}. \label{eq:bcom}
\end{equation}
It is probable that the action of $\hat{\theta}_{\pm}$ on the
wavefunction $\chi$ is given by multiplication
\begin{equation}
\hat{\theta}_{\pm}\chi(\theta_{+},\theta_{-})=
\theta_{\pm}\cdot\chi(\theta_{+},\theta_{-}) \quad.
\end{equation}
To determine the action of $\hat{r}_{\pm}$, however, we have to
know the integration measure or the inner product, the
determination of which in turn requires the knowledge of
transformation properties of the wavefunction and of the basic
operators under the modular group $\Gamma=PSL(2,{\bf Z})$.
Transformations of the classical variables under the
two elementary modular transformations:
$$
S:(\alpha,\beta)\rightarrow(-\beta,\alpha),\quad
T:(\alpha,\beta)\rightarrow(\alpha+\beta,\beta),
$$
prove to be given by the following simultaneous transformations
\begin{eqnarray}
S &:&(\theta_{\pm},r_{\pm})\rightarrow
(\theta_{\pm}+\frac{\pi}{2}, r_{\pm}),\nonumber \\
T &:&(\theta_{\pm},r_{\pm})\rightarrow\left(
\frac{1}{2i}\ln\left\{\frac{e^{i\theta_{\pm}}+\sin\theta_{\pm}}
{e^{-i\theta_{\pm}}+\sin\theta_{\pm}}\right\},(1+\sin2\theta_{\pm}
+\sin^{2}\theta_{\pm})r_{\pm}\right).\label{eq:modul}
\end{eqnarray}
We can show that these transformations preserve the symplectic
structure of ${\cal M}^{\prime}$ and the cotangent bundle structure
of ${\cal M}^{\circ}$. We therefore expect
that under the assumption
made above a consistent quantum theory can be defined on the
\lq\lq fundamental region" ${\cal B}/\Gamma$.

Since the quantum theory constructed as above is defined on $
{\cal M}^{\circ}$ which is larger than the ADM phase space, we may
find a process which is not expected by quantizing the ADM
formalism.
In our quantum theory, momentum eigenstates would play an important
role because each sector is identified by the signature of
$(r_{+},r_{-})$.
Whether our quantization prescription works well, however, depends
on the validity of the two assumptions. First we have assumed that
the relevant gauge
group is the universal covering $\widetilde{SO}(2,2)$ of the
anti-de Sitter group. Second we have used the trick to give the
cotangent bundle structure to the phase space. If either of these
two turns out to be physically irrelevant, such a quantization
prescription cannot be used and we are obliged to take the
geometric quantization methods.

Complete formulation of our quantum theory \cite{ezawa2} includes
a definition of the measure and the quantum version of the modular
transformation (\ref{eq:modul}), its quantum relation to the ADM
formalism \cite{soda}\cite{ezawa}, and its relations to the quantum
theory of refs.\cite{nelson}\cite{CN} and to the geometric
quantization \cite{wood}.

\vskip2.5cm

\noindent Acknowledgments

I would like to thank Prof. K. Kikkawa, Prof. H. Itoyama and H.
Kunitomo for helpful discussions and
careful readings of the manuscript.


\end{document}